# Point group symmetry of polyhedral diagrams in graphic statics

Yefan ZHI[a], Yao LU[b,*], Masoud AKBARZADEH[a,c]

[a] Polyhedral Structures Laboratory, School of Design, University of Pennsylvania, Philadelphia, USA
Pennovation Center, 3401 Grays Ferry Ave. Philadelphia, PA 19146, USA
*Yao.Lu@Jefferson.edu

[b] College of Architecture, Design & Engineering, Thomas Jefferson University, USA
[c] General Robotic, Automation, Sensing and Perception (GRASP) Lab, School of Engineering and Applied Science, University of Pennsylvania, Philadelphia, USA

## Abstract

Symmetry is an implicit objective in structural form-finding that often reconciles efficiency and aesthetics. This paper identifies the symmetry of polyhedral diagrams in three-dimensional graphic statics (3DGS) as point groups and formulates them as constraints, enabling the optimization and manipulation of polyhedral diagrams that preserve such symmetry. 3DGS has been an efficient and effective tool for the form-finding of funicular structures. However, when modifying complex diagrams for design exploration or optimization, one can easily break the symmetry of the reciprocal design input, rendering the result undesirable for practical use. To address this problem, this paper investigates symmetry transformations and introduces point groups, an abstract algebra tool commonly used in crystallography to represent the symmetry and equivalence between a network of atoms (points with labels). It then discusses the hierarchy of symmetry in the geometry types of a polyhedral diagram, and proposes the constraint of symmetry through edge lengths. Based on the crystal symmetry search algorithm by spglib and pymatgen, a fast fingerprinting algorithm is developed to identify the point group of a polyhedral diagram and sort equivalent edges into sets. Finally, the paper shows that the necessary and sufficient condition for preserving the point group symmetry is that each set of edges has the same length. This constraint is compatible with the algebraic formulation of 3DGS and effectively preserves symmetry while reducing the dimension of the solution space. The method is implemented in the PolyFrame 2 plug-in for Rhino and Grasshopper.

## 1. Introduction

Symmetry is a concept that has been widely observed in natural and human-made structures. With its appealing aesthetics, functional advantages, and manufacturing efficiency, symmetry has been extensively studied by both scientists and designers [1]. Their efforts include theorizing and classifying symmetry, identifying it in existing objects, and preserving or creating it in the design process.

In structural form-finding, boundary conditions and design spaces are often symmetric, which naturally leads to the efficient structural forms being symmetric. However, preservation or creation of such symmetry is rarely included in the objective function of structural form-finding. Even when starting from a symmetric input, structural form-finding methods, especially those utilizing a heuristic or stochastic algorithm, can easily break the symmetry. This paper is an effort to reinstate such an implicit constraint in advanced structural form-finding methods, specifically in the algebraic formulation of three-dimensional graphic statics (3DGS).

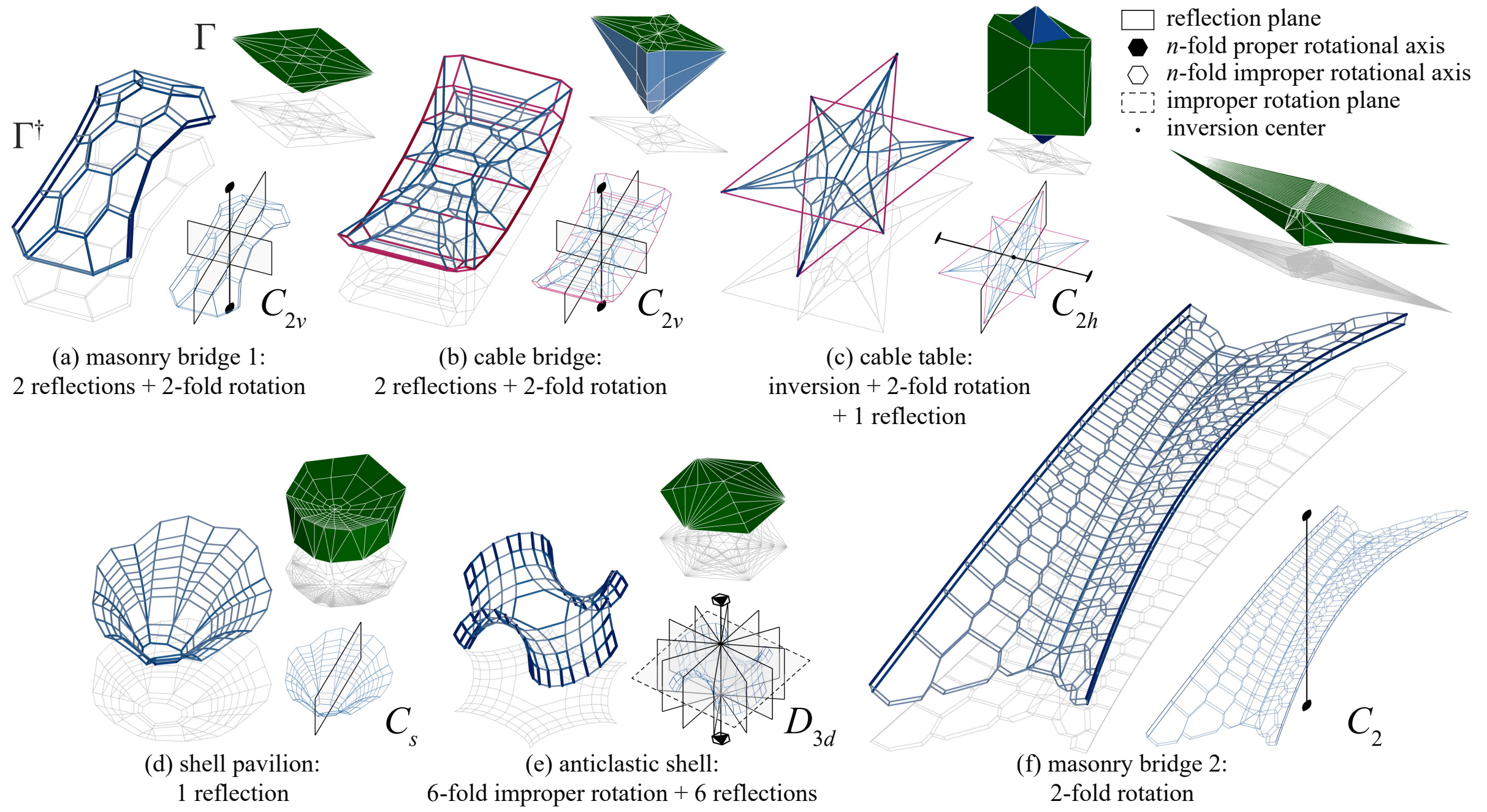

Figure 1: Symmetric forms found using 3DGS, including force ($\Gamma$) and form ($\Gamma^{\dagger}$) diagrams, their symmetry operations and point groups (Schoenflies notation, see Table 2): (a) masonry bridge 1 [5]; (b) cable bridge [6]; (c) cable table [7]; (d) shell pavilion [8]; (e) anticlastic shell [9]; and (f) masonry bridge 2 [10].

### 1.1. Three-dimensional graphic statics

Three-dimensional graphic statics (3DGS) is an intuitive tool for structural form-finding of polyhedral frames, where a *form diagram* represents the location and length of its linear members, and a reciprocal *force diagram* represents the external loads and internal forces [2]. Designers typically start with a customized force diagram which implies equilibrium, and solve for the reciprocal form diagram to create efficient funicular structures whose members are in pure compression or tension. Two solvers are commonly used: the conventional iterative method [3] that perpendicularizes a dual graph to get the reciprocal force diagram, and the algebraic formulation [4] which solves reciprocity as a matrix equation with the edge lengths as variables. For symmetric boundary conditions, if a symmetric form is desired, the design process will start with a symmetric force diagram, which produces a symmetric form diagram through both solvers. Figure 1 gives some examples of symmetric forms found by 3DGS, showing a variety of symmetry operations.

The algebraic formulation of 3DGS establishes the valid polyhedral diagram by solving the vector of all internal edge lengths $\mathbf{q} = \left(q_1, q_2, \ldots, q_{e_{\text{int}}}\right)$ from a linear closing equation $\mathbf{Aq} = \mathbf{0}$ [4]. Algebraic 3DGS is advantageous for its efficiency and accuracy in solving the reciprocal diagrams, as well as the flexibility in design exploration offered by a variety of compatible constraints [11], [12]. When vertex and edge constraints are introduced, the equation is extended as the constrained linear closing equation $\mathbf{Mq} = \mathbf{t}$ [12]. Such formulation also allows linear optimization over the edge lengths towards objectives for construction constraints or benefits [13]. However, in such constrained manipulation and optimization, the symmetry of the force diagram can be easily lost, rendering the result undesirable for practical use. Figure 2 illustrates this problem and outlines the proposed method to address it by embedding the symmetry constraint in the algebraic formulation before further modifications.

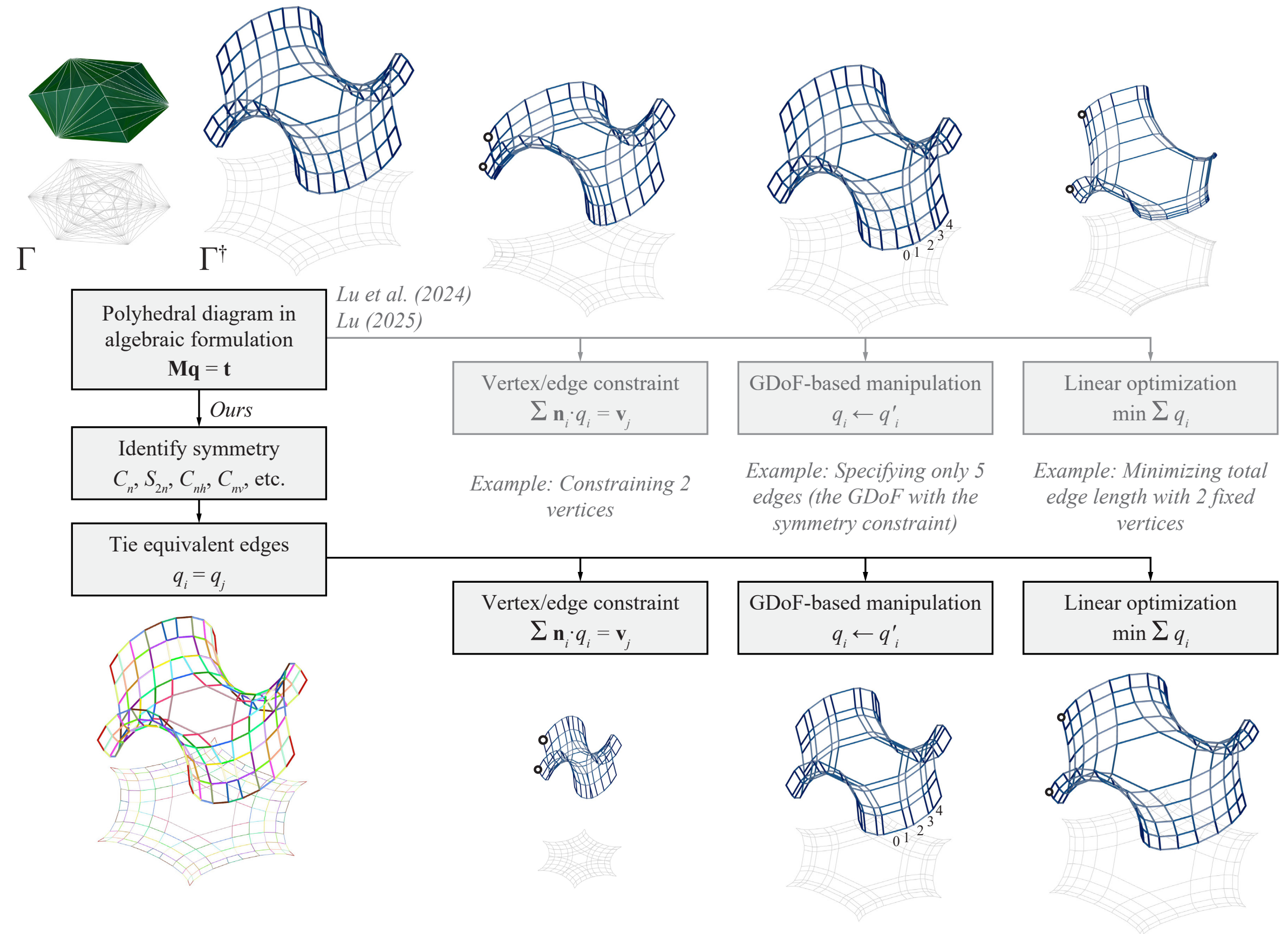

Figure 2: The proposed workflow of adding the symmetry constraint to a polyhedral diagram in its algebraic formulation before further modifications, using an anticlastic shell as an example. Unlike in [12] and [13] where the symmetry is easily lost after optimization, the proposed method preserves the symmetry of the polyhedral diagram.

Table 1: Table of symmetry operations

| **Symbol** | **Operation** | **Elements** | **Parameters** (point, vector, number) |
|---|---|---|---|
| $E$ | identity | – | – |
| $\sigma$ | reflection | • symmetry plane | $O, e$ |
| $i$ | inversion | • inversion center | $O$ |
| $C_n^m$ | proper rotation | • axis<br>• angle | $O, e, \theta$ |
| $S_n^m$ | improper rotation (rotation + reflection) | • axis<br>• angle<br>• symmetry plane | $O, e, \theta$ |

*Note*: $C_n^m$ means a rotation of $360°/n$ performed $m$ times, with a total angle of $\theta = (360° \times m/n)$.

## 2. Point group symmetry

### 2.1. Symmetry operations

A symmetry operation $A$ is a Euclidean transformation that maps a geometry to itself, meaning the geometry is identical after the transformation (formalized in Appendix A). Table 1 summarizes all five types of symmetry operations of a *finite* geometry and the parameters that define them (the parameters are not required by the mathematical definitions of the planes and axes, but are introduced in this paper for ease of discussion). For a finite geometry, if it holds any non-trivial symmetry operation, the *geometric center* (centroid) will stay fixed under the transformation, meaning the geometric center is on all axes and planes, and coincides with the inversion center. In the scope of this paper, we thus define that all $O$ in Table 1 are just the geometric center, and hereby $O$ is known as the *symmetry center*. (Identity is a *trivial* symmetry operation and does not have a symmetry center. It is always "preserved" and will thus not be discussed in this paper's context.)

Table 2: 14 types of three-dimensional point groups

| | |
|---|---|
| Axial groups | $C_n$, $S_{2n}$, $C_{nh}$, $C_{nv}$, $D_n$, $D_{nd}$, $D_{nh}$ |
| Polyhedral groups | $T$, $T_d$, $T_h$, $O$, $O_h$, $I$, $I_h$ |

*Note*: In Schoenflies notation, $n$ is the order of the principal rotation, and $h$, $v$, and $d$ stand for horizontal, vertical, and dihedral reflection planes, respectively.

### 2.2. Point groups

For a geometry to hold multiple distinct symmetry operations, they have to be compatible with each other. More precisely, the composition of any two symmetry operations must also be a symmetry operation of the geometry. The set of all valid symmetry operations of geometry $\Gamma$ forms a *group* $G(\Gamma) = \{A_1, A_2, ...\}$ in the abstract algebra sense, known as the *symmetry group*. Symmetry groups for finite geometries are called *point groups*, emphasizing the existence of a unique symmetry center for non-trivial symmetry operations. Point groups are commonly used in fields such as molecular chemistry and crystallography, where they assist the determination and classification of materials' properties and behaviors [14]. Table 2 summarizes all 14 types of three-dimensional point groups.

Schoenflies notation is adopted in this paper for names of symmetry operations and point groups. (It is beyond the scope of the paper to introduce the Schoenflies notation and the effective expression of symmetry operations as point groups. For more details, see [15], [16], [17].) The name of the point group, along with the parameters (see Table 1), fully describes all symmetry operations of the geometry, thereby suggesting equivalence between parts of the geometry (e.g. atoms) and similarity between different geometries/structures. For example, in Figure 1a and b, the two bridges share the same point group of $C_{2v}$, meaning they are "equally" symmetric, and for each bridge, the four corners are "equivalent" to each other.

Some current literature has explored the application of symmetry groups in architectural design [18], mechanical design [19], and structural design [20], [21]. However, they are mostly confined to 2D and do not provide practical methods to utilize 3D point group symmetries. This paper offers a method to identify and constrain the 3D point group symmetries in algebraic graphic statics, and can be expanded to polyhedral systems in general.

## 3. Identifying symmetry of polyhedral diagrams

### 3.1. Geometry types

A 3D polyhedral diagram $\Gamma = (V, E, F, C)$ is a network of four geometry types: from low-level to high-level, vertices $V$, edges $E$, faces $F$, and cells $C$. It is a collection of elements with distinct geometry types, and thus cannot be represented as a simple subset of the metric space. The definition of symmetry operations has to be expanded to include geometric elements with different types (In Appendix A, Definition 1 and Definition 2), and symmetry can be discussed only if the equivalence relation between the elements can be evaluated (only practical in Definition 2).

In a polyhedral diagram, lower-level elements are not allowed to exist if they do not form some higher-level elements. Different symmetry might be observed in different geometry types. Symmetry operations of a higher-level geometry type (e.g., edges) are also observed in the lower-level geometry type that forms it (e.g., vertices), since all vertices are just the ends of all edges and will inherit the symmetry operations of edges, $A \in G(E) \Rightarrow A \in G(V)$. However, the opposite is not necessarily true, since a vertex symmetry operation would create symmetric vertex pairs that may or may not be connected by an edge, where the symmetry can be lost in the edge level. Therefore, $G(E)$ is a subgroup of $G(V)$, $G(E) \leq G(V)$, meaning all symmetry operations in $G(E)$ are also in $G(V)$, and it is possible that $G(E) < G(V)$. This relation extends to all four geometry types:

$$G(C) \leq G(F) \leq G(E) \leq G(V). \tag{1}$$

*An exact algorithm?* The symmetry of a collection of points can be identified efficiently by a variety of methods. For edges, faces, and cells, as suggested by Definition 2, the *exact* algorithm has to evaluate the equivalence relation between these elements, which is only practical through comparing the lower-level elements that form them. The combinatorial nature of this process makes it unwise to pursue an exact algorithm (see discussion in [22]), especially when a polyhedral diagram can easily have hundreds of vertices, edges, faces, and cells.

### 3.2. Identifying symmetry of the edges

Since the algebraic form is essentially the vector of all internal edge lengths $\mathbf{q} = \left(q_1, q_2, ..., q_{e_{\text{int}}}\right)$, we assume that $G(C) = G(F) = G(E)$ and focus on the edge symmetry. For non-complex faces, edges are non-intersecting, making their midpoints ideal as representative points for equivalence evaluation. Therefore, we propose using a fingerprinting algorithm that identifies $G(E)$ using their midpoints. The midpoints alone are not accurate enough. (For instance, the midpoint of a diagonal line of a rectangle is always the centroid of the rectangle, while the centroid of the rectangle can be on either of the diagonal lines, thereby not reflecting the difference between two different scenarios.) A tag for each edge should be included in the fingerprints. The tag should be based on invariant properties, which could include the length of the edge, the degree (valency) of the incident vertices, the length of incident edges, the degree of incident faces, etc. In practice, to get an integer-based tag for an edge $(u, v)$, we use the length of the edge normalized by a reference length and rounded $\left[\ell_{(u,v)}/\ell_{\text{ref}}\right]$, and the degree of the incident vertices $\deg(u)$, $\deg(v)$ (in no particular order since they are arbitrarily ordered). The fingerprint of the edge can be a weighted sum hash

$$H(u, v) = \left(P_1 \left[\frac{\ell_{(u,v)}}{\ell_{\text{ref}}}\right] + P_2 \ \min\{\deg(u), \deg(v)\} + P_3 \ \max\{\deg(u), \deg(v)\}\right) \bmod P_0 \tag{2}$$

where $P_1$, $P_2$, $P_3$ are prime weights, $P_0$ is a large prime number, and the reference length can be set as a fraction of the minimum edge length $\ell_{\text{ref}} = f_{\text{ref}} \min_{(u,v)\in E} \ell_{(u,v)}$. Any computation library that computes the symmetry of a collection of points with integer tags can then be used to identify the edge symmetry.

*Implementation.* We utilize the Python library for material analysis *pymatgen* [23] whose symmetry module is based on the crystal symmetry library *spglib* [24]. Pymatgen offers an iterative algorithm for identifying point/space group types, while allowing arbitrary initial orientation of the structure and a certain amount of distortion of the point coordinates. The interface allows tagging vertices as atoms of the periodic table, with an atomic number between 1 and 118. Therefore, in (Equation 2) we adopt $P_0 = 113$, and $P_1 = 1$, $P_2 = 11$, $P_3 = 17$ (the normalized length is generally bigger than the degrees of the vertices). The factor of reference length is set as $f_{\text{ref}} = 0.0185$.

## 4. Preserving symmetry in polyhedral diagrams

If there is a symmetry operation that maps one element to another, they are *equivalent* elements. The point group analysis thus divides all edges into equivalent sets. Appendix B proves that the necessary and sufficient condition for a connected polyhedral diagram in its algebraic formulation to still hold all edge symmetry after the edge lengths are changed from $\mathbf{q}$ to $\mathbf{q}'$ is that all edges in an equivalent set still have the same length. The constraint to tie the lengths of edges can be formulated as linear equations compatible with the algebraic formulation. For a group of $k$ equivalent edges with lengths $q_{i_1}, q_{i_2}, ..., q_{i_k}$, the constraint can be formulated as $k-1$ linear equations between pairs

$$\left(q_{i_j}, q_{i_{j+1}}\right), \quad 1 \leq j \leq k-1. \tag{3}$$

To tie the lengths of a pair $(q_l, q_r)$ we require $q_l - q_r = 0$. Let $\mathbf{e}_i = (0, 0, ..., 0, 1, 0, ..., 0)$ be the $e_{\text{int}}$-tuple with all components being 0 except the $i$-th component being 1, the constraint can be written as

$$(\mathbf{e}_l - \mathbf{e}_r) \cdot \mathbf{q} = \mathbf{0}. \tag{4}$$

All equivalent sets can be assembled to form a symmetry constraint matrix

$$\mathbf{S} = \begin{pmatrix} \left(\mathbf{e}_{l_1} - \mathbf{e}_{r_1}\right)^\top \\ \left(\mathbf{e}_{l_2} - \mathbf{e}_{r_2}\right)^\top \\ \vdots \end{pmatrix} \tag{5}$$

and the symmetry constraint is

$$\mathbf{S}\mathbf{q} = \mathbf{0}. \tag{6}$$

The closing matrix $\mathbf{M}$ in the constrained linear closing equation $\mathbf{M}\mathbf{q} = \mathbf{t}$ is thus replaced by

$$\mathbf{M}_{\text{sym}} = \begin{pmatrix} \mathbf{M} \\ \mathbf{S} \end{pmatrix} \tag{7}$$

to include the symmetry constraint.

## 5. Results and applications

The geometric degrees of freedom (GDoF, denoted $m$) of the polyhedral diagram is the dimension of the solution space of the closing equation. It means there exists a set of $m$ edges whose lengths can be freely specified to produce a unique solution of $\mathbf{q}$. The set can be determined through the reduced row echelon form (RREF) of $\mathbf{M}$ or $\mathbf{M}_{\text{sym}}$ [4]. For the case studies in Figures 1 and 3, the symmetry constraint significantly reduces the GDoF (Table 3), making the design space more manageable for heuristic design exploration and optimization.

Table 3: Symmetry identification and constraint results from Figures 1 and 3

| | **Example** | **Group** $G$ | **Order** $\lvert G\rvert$ | $\frac{m_{\text{raw}}}{m_{\text{sym}}}$ | **#Rows of S** | $e_{\text{int}}$ | $m_{\text{raw}}$ | $m_{\text{sym}}$ | **Time** |
|---|---|---|---|---|---|---|---|---|---|
| (a) | masonry bridge 1 | $C_{2v}$ | 4 | 2.14 | 76 | 108 | 15 | 7 | 0.03s |
| (b) | cable bridge 1 | $C_{2v}$ | 4 | 1.82 | 157 | 222 | 20 | 11 | 0.06s |
| (c) | cable table | $C_{2h}$ | 4 | 3 | 78 | 110 | 12 | 4 | 0.03s |
| (d) | shell pavilion | $C_s$ | 2 | 1.29 | 154 | 322 | 9 | 7 | 0.08s |
| (e) | anticlastic shell | $D_{3d}$ | 12 | 5.6 | 245 | 270 | 28 | 5 | 0.12s |
| (f) | masonry bridge 2 | $C_2$ | 2 | 1.96 | 485 | 971 | 102 | 52 | 4.93s |
| (g) | 2D subdivision | $C_{2v}$ | 4 | 1.75 | 10 | 21 | 7 | 4 | 0.01s |
| (h) | funnel shell | $C_{4v}$ | 8 | 3.71 | 126 | 148 | 26 | 7 | 0.06s |
| (i) | cellular solid | $O_h$ | 48 | 14.7 | 466 | 480 | 103 | 7 | 0.47s |

The order of the point group $|G|$ is the number of symmetry operations in the group, and thus an upper bound of the size of each equivalent set. The reduction factor of $m$ is no greater than $|G|$,

$$1 \leq m_{\text{raw}}/m_{\text{sym}} \leq |G|, \tag{8}$$

because the $m_{\text{raw}}$ independent edges, when sorted into equivalent sets, will exist in at least $m_{\text{raw}}/|G|$ sets according to the pigeonhole principle, providing a lower bound for $m_{\text{sym}}$. (It is possible that $m_{\text{raw}}/m_{\text{sym}} = |G|$. For instance, an inversion group with two tetrahedra meeting at the inversion center has $m_{\text{raw}} = 2$, $m_{\text{sym}} = 1$, and $|G| = 2$.) This observation is in line with the experimental results in Table 3.

*Correctness.* While the fingerprint algorithm is not exact, in our experiments it has always correctly identified the symmetry of the edges and sorted them into equivalent sets. The equivalent edges are colored the same (e.g. Figure 2) for visual inspection. Misidentification is easy to be spot by the user and in such rare cases, manual correction is needed.

*Software implementation and computation time.* The proposed method is implemented in the PolyFrame 2 plug-in [25] for Rhino and Grasshopper. In the fingerprint (Equation 2), the degree can be computed in constant time for each vertex with the edge topology saved as an adjacency list, therefore the bottleneck of the computation time is the external library used. Table 3 shows that the computation time is impacted by the order of the point group $|G|$ and the number of edges $e_{\text{int}}$, but is generally negligible.

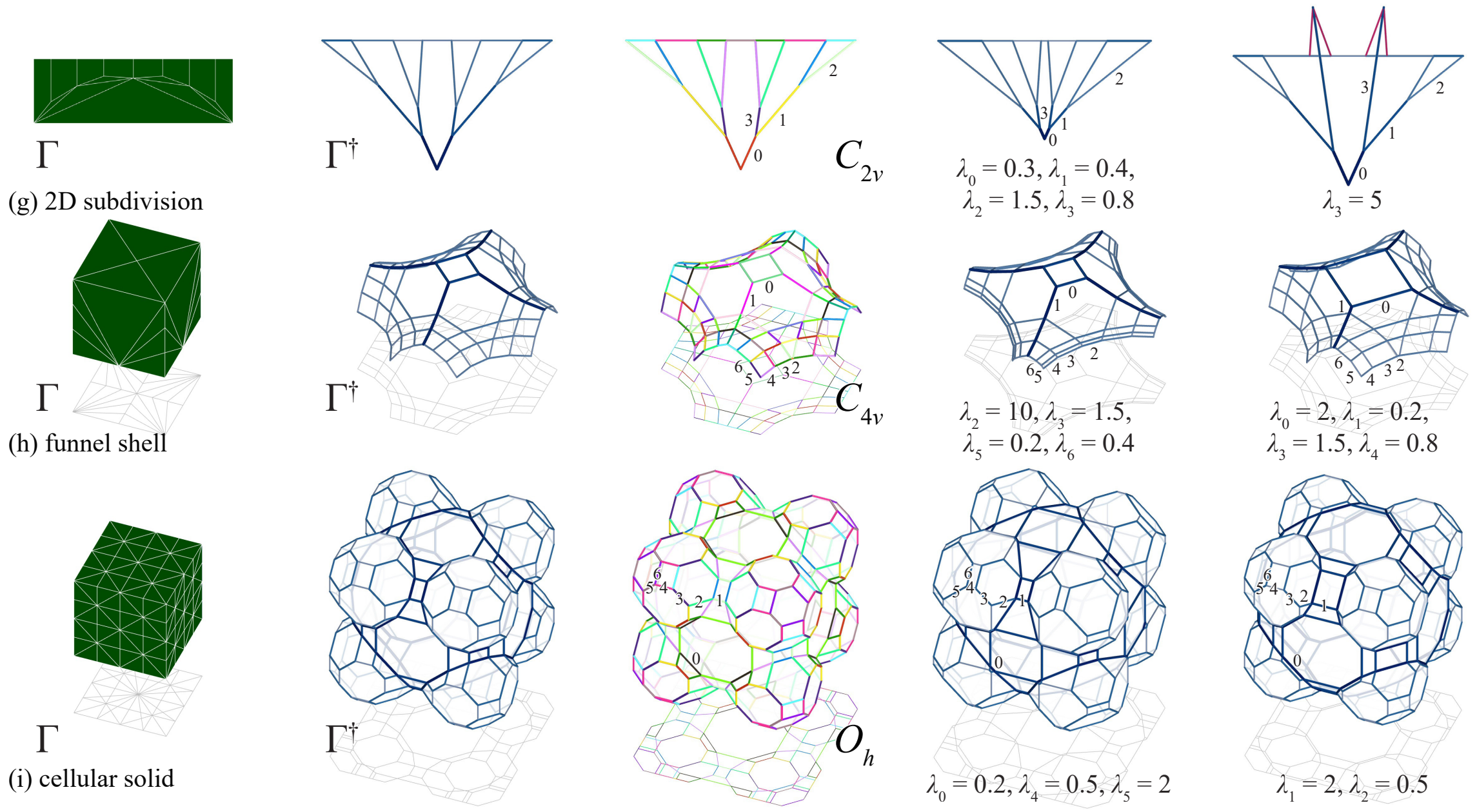

Figure 3: GDoF-based manipulation examples, from [26], showing point groups and equivalent sets (color-coded per set), independent edges with symmetry constraints (labeled), and scaling factors ($\lambda_i$): (g) 2D subdivision (using 3D notation); (h) funnel shell; and (i) cellular solid.

*Example of application: GDoF-based manipulation.* The proposed symmetry constraint is a component compatible with any algebraic 3DGS form-finding frameworks. One example would be modifying the diagram by specifying the lengths of the independent edges. Figure 3 shows the effect of modifying edge lengths of an existing force diagram $\Gamma^\dagger$ with scaling factors, $q_i \leftarrow \lambda_i q_i$. With the reduced GDoF, such manipulation becomes manageable by hand, allowing efficient heuristic explorations.

## 6. Conclusion

This paper presented a fast symmetry identification and preservation method for algebraic 3DGS through fingerprinting and linear equations. The following contributions are highlighted: (1) The concise and fully automated component is applicable in any algebraic 2D/3D graphic statics frameworks and can be combined with other constraints in a flexible order. (2) The significant reduction of the dimension of the solution space makes any optimization and manipulation methods more manageable and efficient. (3) Ready-to-use software implementation is provided.

Future work includes methods involving subgroups or partial symmetry (Figure 3i is an example where it is lost), and the potential to restore symmetry heuristically in a distorted diagram modeled by mistake.

## Appendix A. Formal definition of a symmetry operation

**Definition 1** (Symmetry of geometry as function): Let an $n$-dimensional geometry be presented by a function $f : \mathbb{R}^n \to S$ where $S$ is any set with distinct identities (e.g., existence, type of geometry, color, material, label, etc.). A Euclidean transformation $A$ is a symmetry operation for this geometry if and only if $f(Ax) = f(x), \forall x \in \mathbb{R}^n$.

**Definition 2** (Symmetry of geometry as set, extension of Definition 1): Let a geometry be presented by a set $S$ whose elements can be transformed as $\mathbb{R}^n$ vectors and might embed any additional properties (e.g., type of geometry, color, material, label, etc.). A Euclidean transformation $A$ is a symmetry operation for this geometry if and only if $A$ is a bijection from $S$ to itself (a permutation of $S$).

## Appendix B. Edge length theorem

**Theorem 1** (Edge length theorem): For a connected polyhedral diagram $\Gamma$ in its algebraic formulation, after the edge lengths are changed from $\mathbf{q}$ to $\mathbf{q}'$ to form $\Gamma'$, the necessary and sufficient condition for $\Gamma'$ to still hold all edge symmetry is that all edges in an equivalent set still have the same length.

*Proof*: Necessity (symmetry $\Rightarrow$ equal length constraint): trivial.

Sufficiency (equal length constraint $\Rightarrow$ symmetry): We prove that any symmetry transformation $A$ of $\Gamma$'s edges $\Gamma.E$ is still preserved in $\Gamma'.E$ through the following three steps:

$$A \in G(\Gamma.E) \Rightarrow A \in G(\Gamma.V) \Rightarrow A \in G(\Gamma'.V) \Rightarrow A \in G(\Gamma'.E) \tag{9}$$

$A \in G(\Gamma.E) \Rightarrow A \in G(\Gamma.V)$ is trivial following the discussion in Section 3.1. Since edge topology is preserved, we also know that $A \in G(\Gamma.E), A \in G(\Gamma'.V) \Rightarrow A \in G(\Gamma'.E)$. It suffices to prove that $A \in G(\Gamma.E), A \in G(\Gamma.V) \Rightarrow A \in G(\Gamma'.V)$. Since identity is trivial, and reflection and inversion can be reduced to improper rotation, it suffices to prove this for proper and improper rotational symmetry.

Any symmetry transformation $A \in G(\Gamma.E) \cap G(\Gamma.V)$ is a permutation (from Definition 2) and divides all vertices into cyclic lists $l_1, l_2, ....$ Let $k$ be the order of $A$ (length of the longest cyclic list). Consider any cyclic list with $k$ vertices $l = (p_1, p_2, ..., p_k)$, we have

$$p_{i+1} = Ap_i, \quad 1 \leq i \leq k. \tag{10}$$

Consider the translation vectors $v_i = p_{i+1} - p_i$. Since $\Gamma$ is connected, there is a path between $p_1$ and $p_2$. Then $v_1$ is the summation of the vectors representing these edges. Since all these edges have corresponding image edges in $\Gamma$ following $A$, these edges form a path between $p_2$ and $p_3$ as $v_2 = p_3 - p_2$, all the way up to $v_k$. According to the edge transformations, $v_{i+1} = Av_i$.

In the new diagram $\Gamma'$, the translation vectors are $v'_i = p'_{i+1} - p'_i$. The directions of all edges are preserved, and edges in an equivalent set still have the same length. We still have $v'_{i+1} = Av'_i$. The cyclic relation:

$$\sum_{i=1}^{k} v'_i = \sum_{i=1}^{k} A^{i-1} v'_1 = 0. \tag{11}$$

Consider $A$ as a proper rotation first. With respect to the rotational axis direction $e$, decompose $v'_1 = v'_\parallel + v'_\perp$. Then $A^i v'_1 = v'_\parallel + A^i v'_\perp$. (Equation 11) becomes

$$kv'_\parallel + \sum_{i=1}^{k} A^{i-1} v'_\perp = 0. \tag{12}$$

Then $v'_\parallel = 0$, $v' = v'_\perp$. (Equation 12) shows that a closed polygon is formed by $v'_1, v'_2, ..., v'_k$. Then $l' = (p'_1, p'_2, ..., p'_k)$ also holds a rotational symmetry $A'$ with the same rotational axis vector $e$ and angle $\theta$:

$$p'_{i+1} = A'p'_i, \quad 1 \leq i \leq k. \tag{13}$$

For any vertex $r'_1 \notin \ell$ in $\Gamma'$, since $\Gamma'$ is connected, there is a path from $p'_1$ to $r'_1$, represented by translation vector $w'_1 = r'_1 - p'_1$. Since the directions of all edges are preserved, and edges in an equivalent set still have the same length, there is also a path from $p'_2$ to a vertex $r'_2$, where $w'_2 = r'_2 - p'_2 = Aw'_1 = A'w'_1$ (for transforming a vector, the origin does not matter, and $A$, $A'$ are equivalent). Therefore,

$$A'r'_1 = A'(p'_1 + w'_1) = A'p'_1 + A'w'_1 = p'_2 + w'_2 = r'_2, \tag{14}$$

meaning there is an image vertex of $r'_1$ following $A'$ in $\Gamma'$. From (Equation 13) and (Equation 14), $A' \in G(\Gamma'.V)$. Since the symmetry center of the entire diagram is always the geometric center, $A'$ and $A$ are identical, $A \in G(\Gamma'.V)$.

The proof for $A$ as an improper rotational symmetry is similar, except that $k$ is an even number, $A^i v'_1 = (-1)^i v'_\parallel + A^i v'_\perp$, and $v'_\parallel$ can have any magnitude. ■

*Note*: While all symmetry would still hold, the point group of $\Gamma'$ might change, as new symmetry operations may be created, where the old point group becomes a subgroup of the new one, $G(\Gamma) \leq G(\Gamma')$. An example would be getting a square from a rectangle.

## Acknowledgements

This research is funded by the Thomas Jefferson University Faculty Seed Grant to Yao Lu and the National Science Foundation Future Eco Manufacturing Research Grant (NSF, FMRGCMMI2037097) to Masoud Akbarzadeh.